\documentclass[preprint]{aastex61}
\usepackage{bm}
\usepackage{graphicx}
\usepackage{color}
\usepackage{natbib}

\newcommand{\beq}{\begin{equation}}
\newcommand{\eeq}{\end{equation}}

\shorttitle{\small Particle Acceleration in Relativistic 3D Pair-Plasma Reconnection}
\shortauthors{Werner \& Uzdensky}

\begin{document}

\title{Nonthermal Particle Acceleration in 3D Relativistic Magnetic Reconnection in Pair Plasma}

\correspondingauthor{Gregory R. Werner}
\email{Greg.Werner@colorado.edu}

\author{Gregory R. Werner} 
\affiliation{Center for Integrated Plasma Studies, Physics Department, \\390 UCB, 
University of Colorado, Boulder, CO 80309, USA}

\author{Dmitri A. Uzdensky} 
\affiliation{Center for Integrated Plasma Studies, Physics Department, \\390 UCB, 
University of Colorado, Boulder, CO 80309, USA}
\affiliation{Institute for Advanced Study, 1 Einstein Dr., Princeton, NJ 08540, USA}



\begin{abstract}

As a fundamental process converting magnetic to plasma energy in high-energy astrophysical plasmas, relativistic magnetic reconnection is a leading explanation for the acceleration of particles to the ultrarelativistic energies necessary to power nonthermal emission (especially X-rays and gamma-rays) in pulsar magnetospheres and pulsar wind nebulae, coronae and jets of accreting black holes, and gamma-ray bursts.  An important objective of plasma astrophysics is therefore the characterization of nonthermal particle acceleration (NTPA) effected by reconnection.  Reconnection-powered NTPA has been demonstrated over a wide range of physical conditions using large two-dimensional (2D) kinetic simulations.  However, its robustness in realistic 3D reconnection---in particular, whether the 3D relativistic drift-kink instability (RDKI) disrupts NTPA---has not been systematically investigated, although pioneering 3D simulations have observed NTPA in isolated cases.  Here we present the first comprehensive study of NTPA in 3D relativistic reconnection in collisionless electron-positron plasmas, characterizing NTPA as the strength of 3D effects is varied systematically via the length in the third dimension and the strength of the guide magnetic field.  We find that, while the RDKI prominently perturbs 3D reconnecting current sheets, it does not suppress particle acceleration, even for zero guide field; fully 3D reconnection robustly and efficiently produces nonthermal power-law particle spectra closely resembling those obtained in~2D. 
This finding provides strong support for reconnection as the key mechanism powering high-energy flares in various astrophysical systems.  We also show that strong guide fields significantly inhibit NTPA, slowing reconnection and limiting the energy available for plasma energization, yielding steeper and shorter power-law spectra.

\end{abstract}

\keywords{ acceleration of particles --- magnetic reconnection ---
relativistic processes --- pulsars: general --- gamma-ray burst: general
--- galaxies: jets}



\section{Introduction}
\label{sec-intro}

A long-standing puzzle in high-energy plasma astrophysics is the mechanism behind nonthermal particle acceleration (NTPA) that produces power-law X-ray and $\gamma$-ray spectra observed in pulsar wind nebulae (PWN), coronae and jets of accreting black holes (BHs) in X-ray Binaries (XRBs) and Active Galactic Nuclei (AGN) including blazars, and Gamma Ray Bursts (GRBs).
A leading candidate is magnetic reconnection, a basic plasma process that rapidly converts magnetic into particle kinetic energy through magnetic field rearrangement and relaxation.
In the high-energy universe, magnetic reconnection is often relativistic: the energy density of the reconnecting magnetic field $B_0$ exceeds that of the ambient plasma (including rest-mass), heating plasma to relativistic temperatures, driving relativistic flows, and accelerating particles to ultrarelativistic energies.  
Relativistic reconnection has been invoked in electron-ion or mixed-composition plasmas in accreting BH coronae \citep{Beloborodov-2017}, GRBs \citep{Drenkhahn_Spruit-2002, Giannios-2008, McKinney_Uzdensky-2012}, and blazar jets \citep{Giannios_etal-2009, Nalewajko_etal-2011}, but also, especially, in electron-positron pair plasmas in pulsar magnetospheres \citep{Lyubarsky-1996, Uzdensky_Spitkovsky-2014, Cerutti_etal-2015}, pulsar winds \citep{Coroniti-1990, Arka_Dubus-2013}, PWN \citep{Uzdensky_etal-2011, Cerutti_etal-2012a, Cerutti_etal-2013,Cerutti_etal-2014b},  and magnetar flares \citep{Lyutikov-2003, Lyutikov-2006, Uzdensky-2011}. 
Beyond its astrophysical applications, relativistic pair reconnection is important as the simplest reconnection scenario, a reference case for studying effects of plasma composition, collisions, radiation, etc.

Recently, first-principles particle-in-cell (PIC) kinetic simulations have significantly advanced understanding of NTPA in relativistic pair-plasma reconnection, mostly in two dimensions (2D) 
\citep{Zenitani_Hoshino-2001, Zenitani_Hoshino-2005, Zenitani_Hoshino-2007, Jaroschek_etal-2004a, Lyubarsky_Liverts-2008, Liu_etal-2011, Bessho_Bhattacharjee-2007, Cerutti_etal-2012b, Cerutti_etal-2013, Sironi_Spitkovsky-2014, Guo_etal-2014, Guo_etal-2015, Liu_etal-2015, Sironi_etal-2015, Sironi_etal-2016, Werner_etal-2016}, with a few three-dimensional (3D) studies \citep{Jaroschek_etal-2004a, Zenitani_Hoshino-2008, Cerutti_etal-2014a, Kagan_etal-2013, Sironi_Spitkovsky-2014, Guo_etal-2014, Guo_etal-2015}.
Such studies are very challenging because even the qualitative character of reconnection dynamics, hence also NTPA, depends on the scale separation between the global  
system size~$L$ and the microscopic plasma scales, e.g., the average electron gyroradius in the reconnection layer, $\bar{\rho}_e = \bar{\gamma} m_e c^2/e B_0$, where $\bar{\gamma} m_e c^2$ is the average relativistic energy of electrons energized by reconnection. 
When $L/\bar{\rho}_e$ is large, as in astrophysical systems, reconnection is highly dynamic, with many secondary magnetic islands (plasmoids, or in~3D, flux ropes) and inter-plasmoid current sheets \citep{Bhattacharjee_etal-2009, Uzdensky_etal-2010, Loureiro_etal-2012}.
The relevant large-system plasmoid-dominated regime, in which reconnection characteristics like the particle energy distribution $f(\gamma)$ no longer depend on~$L$, is realized above a critical size $L_c \sim 10^2\bar{\rho}_e$ \citep{Werner_etal-2016}.

Only recently have 2D PIC simulations managed to probe this large-system regime systematically, confirming reconnection-powered NTPA \citep[e.g.,][]{Sironi_Spitkovsky-2014, Guo_etal-2014, Guo_etal-2015, Sironi_etal-2015, Werner_etal-2016}. 
Importantly, they mapped out key acceleration parameters versus~$L$ and the ambient plasma and magnetic field conditions, which are characterized by ``cold'' and ``hot'' plasma magnetizations, $\sigma\equiv B_0^2/4 \pi n_b m_e c^2$ and $\sigma_{h} \equiv B_0^2/4 \pi h$ \cite[see, e.g.,][]{Melzani_etal-2014b}, where $B_0$ is the reconnecting magnetic field, 
$n_b$ is the upstream (background) particle density, and $h$ is the upstream relativistic plasma enthalpy density 
including the rest-mass contribution~$n_b m_e c^2$. 
Thus the particle energy power-law index, $p(L,\sigma_h) \equiv -d\ln f/d\ln\gamma$, was found to become independent of $L$ as $L\rightarrow\infty$, approaching a value consistent with~1 in the ultrarelativistic limit $\sigma_h \simeq \sigma \gg 1$ \citep[in 2D simulations with no guide field,][]{Guo_etal-2014,Werner_etal-2016}.
In addition, \cite{Werner_etal-2016} investigated the high-energy cutoff $\gamma_c(L,\sigma)$ of $f(\gamma)$: for large systems, $\gamma_{c}\sim 4\sigma$ 
(independent of~$L$), while for small systems, $\gamma_{c} \sim 0.1 L/\rho_0$, where $\rho_0\equiv m_e c^2/eB_0$. 
These empirical results thus identified the above-mentioned critical size $L_c\simeq 40 \sigma \rho_0$, where the two cutoffs meet, beyond which the NTPA properties become insensitive to~$L$. 


However, these large-scale 2D studies naturally cannot describe the fundamentally-3D nature of reconnection dynamics, including the rapid development of the relativistic drift-kink instability (RDKI), whose effect on NTPA is still debated.
Whereas \cite{Zenitani_Hoshino-2008} claimed that RDKI suppresses particle acceleration in~3D, recent studies \citep{Sironi_Spitkovsky-2014, Guo_etal-2014} suggest that RDKI eventually saturates, restoring (2D-like) NTPA. 
If RDKI suppresses NTPA, a guide magnetic field (which suppresses RDKI) may bolster particle acceleration~\citep{Zenitani_Hoshino-2008}; this issue, however, remains unsettled. 

Resolving these important issues requires a thorough, systematic study characterizing 3D effects on NTPA using large simulations.
To accomplish this, we perform a set of 3D PIC simulations with varying box aspect ratio $L_z/L_x$ and guide magnetic field~$B_{gz}$, which govern the three-dimensionality of the reconnection process (e.g., small $L_z$ or large $B_{gz}$ suppress RDKI).
Focusing on particle energy spectra, we then present an unambiguous demonstration of NTPA in relativistic pair-plasma reconnection,
comprehensively characterizing NTPA from 2D to~3D.
Specifically, we describe the power-law index $p$ as a function of $L_z/L_x$ and~$B_{gz}$, showing that: 
(1) despite RDKI, 3D reconnection drives NTPA as efficiently as 2D reconnection, yielding remarkably similar spectra;
(2) modest guide fields $B_{gz} \lesssim B_0/4$ barely affect NTPA, but strong $B_{gz} \gtrsim B_0$ dramatically suppress NTPA, yielding steeper spectra.
Our results can provide useful prescriptions for particle spectra produced by reconnection for comparison with astrophysical observations. 


\section{Simulations}
\label{sec-simulations}

Using the electromagnetic PIC code Zeltron \citep{Cerutti_etal-2013},
we run simulations in a box of size $L_x \times L_y \times L_z$ with periodic boundary conditions;
here, $x$ is the direction of reconnecting magnetic field, $y$ is perpendicular to the current layers, and the third dimension $z$ parallels the initial current and guide field.
The 3D set-up extends the 2D double-periodic systems used in our previous paper \citep{Werner_etal-2017submitted} uniformly in~$z$.

The initial state is described by the reversing magnetic field~$B_x$ and background ultrarelativistic Maxwell-J\"{u}ttner pair plasma with temperature $\theta_b=kT_b/m_e c^2 = 275 $ and combined electron and positron density~$n_b$.
Conforming to periodic boundary conditions, the magnetic field undergoes two reversals (supported by two initial current layers of half-thickness $\delta$), $B_x(y) = \pm B_0 \textrm{tanh}[(y-y_{1,2})/\delta]$,
where $y_1=L_y/4$ and $y_2=3L_y/4$ are the midplanes of the two layers.
A uniform guide field $B_{gz}$ is added in the $z$-direction;
we explore $B_{gz}/B_0 \in \{0, 0.25, 0.5, 1, 2\}$.

To study magnetically-dominated reconnection, we initialize the upstream plasma with the largest ``hot'' magnetization resolving the Debye length (described below), $\sigma_h \equiv B_0^2/(4\pi h) = 25$ 
\citep[where $h=4 n_b \theta_b m_e c^2$ in the relativistic limit, cf.][]{Melzani_etal-2014b}.
The ``cold'' magnetization, which sets the overall simulation scale and could be anything consistent with $\sigma_h=25$ and $\theta_b \gg 1$, is $\sigma \equiv B_0^2/(4\pi n_b m_e c^2) = 4\theta_b \sigma_h= 2.75\times 10^4$, as in \citet{Werner_etal-2017submitted}. 

The two current layers are relativistic Harris sheets \citep{Kirk_Skjaeraasen-2003} with simulation-frame density $n_d(y)=n_{d0}\cosh^{-2}[(y-y_{1,2})/\delta]$.
Within each sheet, electrons and positrons drift with opposite average velocities $\pm \beta_{\rm drift} c \hat{\bf z}$ to generate the current supporting the magnetic field reversal, i.e.,
satisfying Ampere's Law with $\delta = B_0 / (4\pi e n_{d0} \beta_{\rm drift})$.
The Harris sheets also provide pressure balance, with rest-frame temperature $\theta_d = (1/2) \gamma_{\rm drift}(n_{b}/n_{d0}) \sigma = 50 \gamma_{\rm drift} (n_{b}/n_{d0}) \theta_b $, where $\gamma_{\rm drift} = (1-\beta_{\rm drift}^2)^{-1/2}$. 
Here we use $\theta_b = 275$, $n_{d0}/n_b = 5$, and $\beta_{\rm drift} = 0.3$, hence $\gamma_{\rm drift} \simeq 1.05$, and $\theta_d = 2890$.

In nonradiative ultrarelativistic pair reconnection, lengths and times scale with~$\sigma$; as long as $\theta_b \gg 1$, simulations with the same $\sigma_h$ but different $\sigma$ are equivalent after rescaling.
Thus, we define a characteristic length scale 
$\rho_c \equiv 2 \sigma \rho_0 = 2 \sigma m_e c^2/eB_0$, 
4 times the gyroradius of a particle with energy $B_0^2/(8\pi n_b)$.%
\footnote{This definition agrees with \citet{Werner_etal-2017submitted}, but differs from $\rho_c\equiv \theta_d \rho_0$ used by \citet{Werner_etal-2016}.
}
The corresponding timescale is $\omega_c\equiv c/\rho_c$.
 Convergence studies led us to choose 40 total macroparticles per cell and grid cell size $\Delta x = \Delta y = \Delta z = \rho_c/24$,
sufficient to resolve both the evolved and initial ($\delta=8\Delta x=\rho_c/3$) current-layer thicknesses, 
and also the upstream Debye length ($\lambda_D = 1.2 \Delta x$), preventing unphysical finite-grid heating \citep{Langdon-1970}.
We use the Courant-Friedrichs-Lewy timestep $\Delta t = 0.99\Delta x/(\sqrt{3}c)$ and typically run until $t \approx 5 L_x/c$ (i.e., $t\approx 200/\omega_c$ for $L_x=40\rho_c$). These parameters yielded better than 1\% energy conservation.

We show results mostly for the fiducial size $L_x=40\rho_c=80\sigma\rho_0$, with additional 
$L_x=20\rho_c$ and~$L_x=64\rho_c$ simulations for $B_{gz}=0$. 
All systems have $L_y =2L_x$ to prevent undesirable interaction between the two layers.
By varying $L_z$ from a single cell (2D) up to $4L_x$, we explore the 3D nature of reconnection.

As in many 2D reconnection studies, the initial magnetic field is perturbed uniformly in~$z$, as described by the vector potential, e.g., for the lower layer
\begin{eqnarray}
  A_z &=& \left[ 1 + 
    0.01 \cos \frac{2\pi x}{L_x} \cos^2 \frac{2\pi (y - y_1)}{L_y}
    \right] 
    B_0 \delta
    \left[ \ln \textrm{cosh} \frac{y_2-y_1}{2\delta}
          -\ln \textrm{cosh} \frac{y-y_1}{\delta}
    \right]
.\end{eqnarray}
In all our 2D simulations, and in 3D simulations with $B_{gz}\geq 0.25 B_0$, the presence of a perturbation does not change the reconnection rate or NTPA.
In perturbationless 3D runs with $B_{gz}=0$, however, particles from the dense initial Harris sheet avoid becoming trapped in flux ropes, and instead spread about the layer, increasing the immediately-upstream density, effectively lowering $\sigma_h$, and significantly slowing reconnection and hindering NTPA.
Leaving details for a future publication, we emphasize that initializing 3D $B_{gz}=0$ simulations with a perturbation makes them more closely resemble (2D and~3D) $B_{gz}=B_0/4$ simulations.


\section{Results}  
\label{sec-results}

Our main findings are:
(1) magnetic energy dissipation and nonthermal particle acceleration (NTPA) are largely unaffected by three-dimensionality, i.e., by $L_z/L_x$;
and (2) a strong guide field reduces the rate and amount of magnetic dissipation and inhibits NTPA.

We illustrate these results with plots showing energy dissipation and final particle energy spectra in simulations with $L_x=40\rho_c = 80\sigma \rho_0$, for $L_z/L_x$ ranging from 0.001 (one cell in $z$) up to 1 (and up to $L_z/L_x=2$ for $B_{gz}=B_0$ and 4 for $B_{gz}=0$), and for $B_{gz}/B_0 \in \{0, 0.25, 0.5, 1, 2\}$, all with $\sigma_h=25$.

\begin{figure}[h]
\begin{center}
\raisebox{8mm}{\includegraphics[width=8.cm,trim=0 0 0 0]{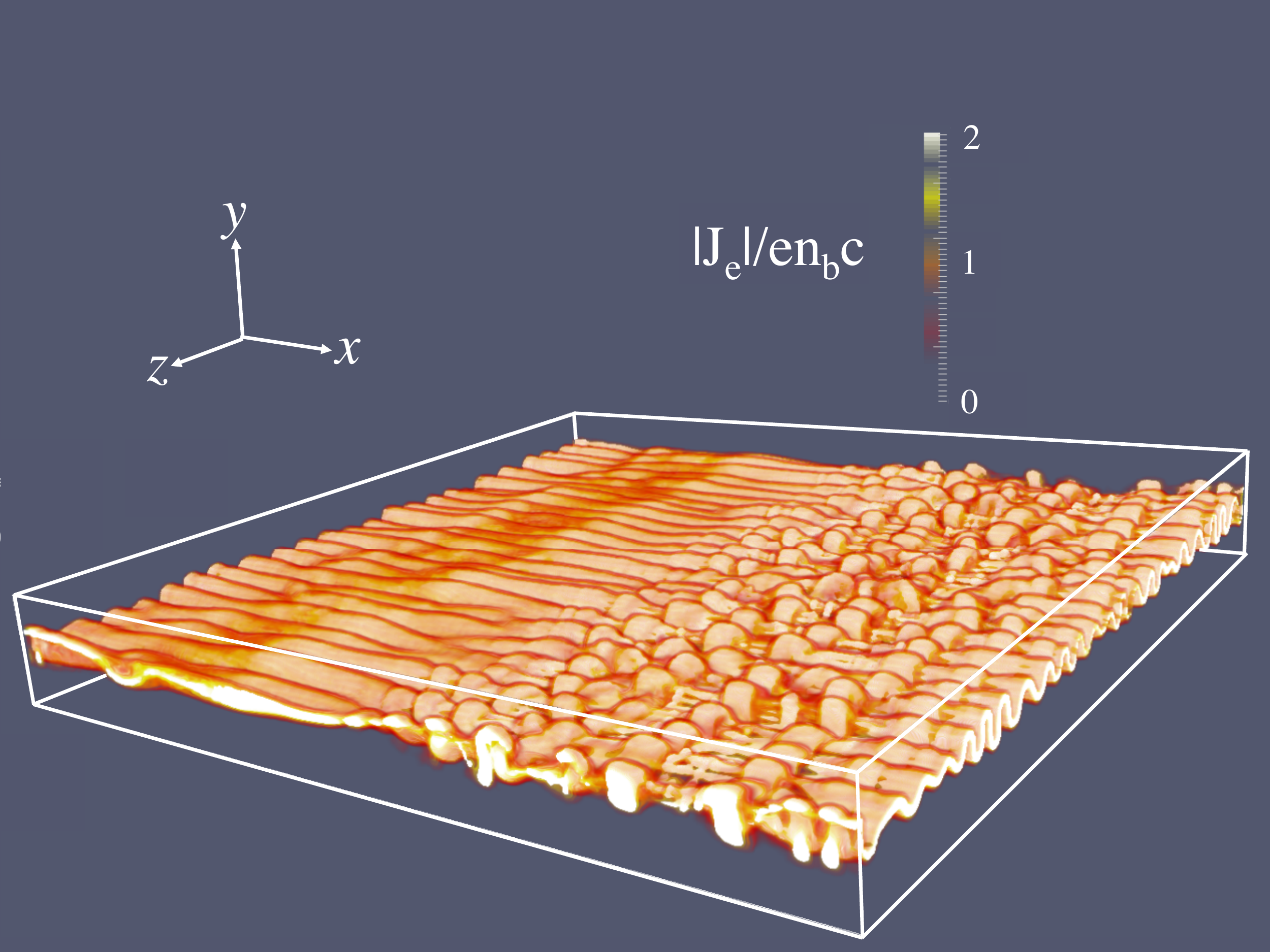}}%
\includegraphics[width=8.cm,trim=0 0 0 0]{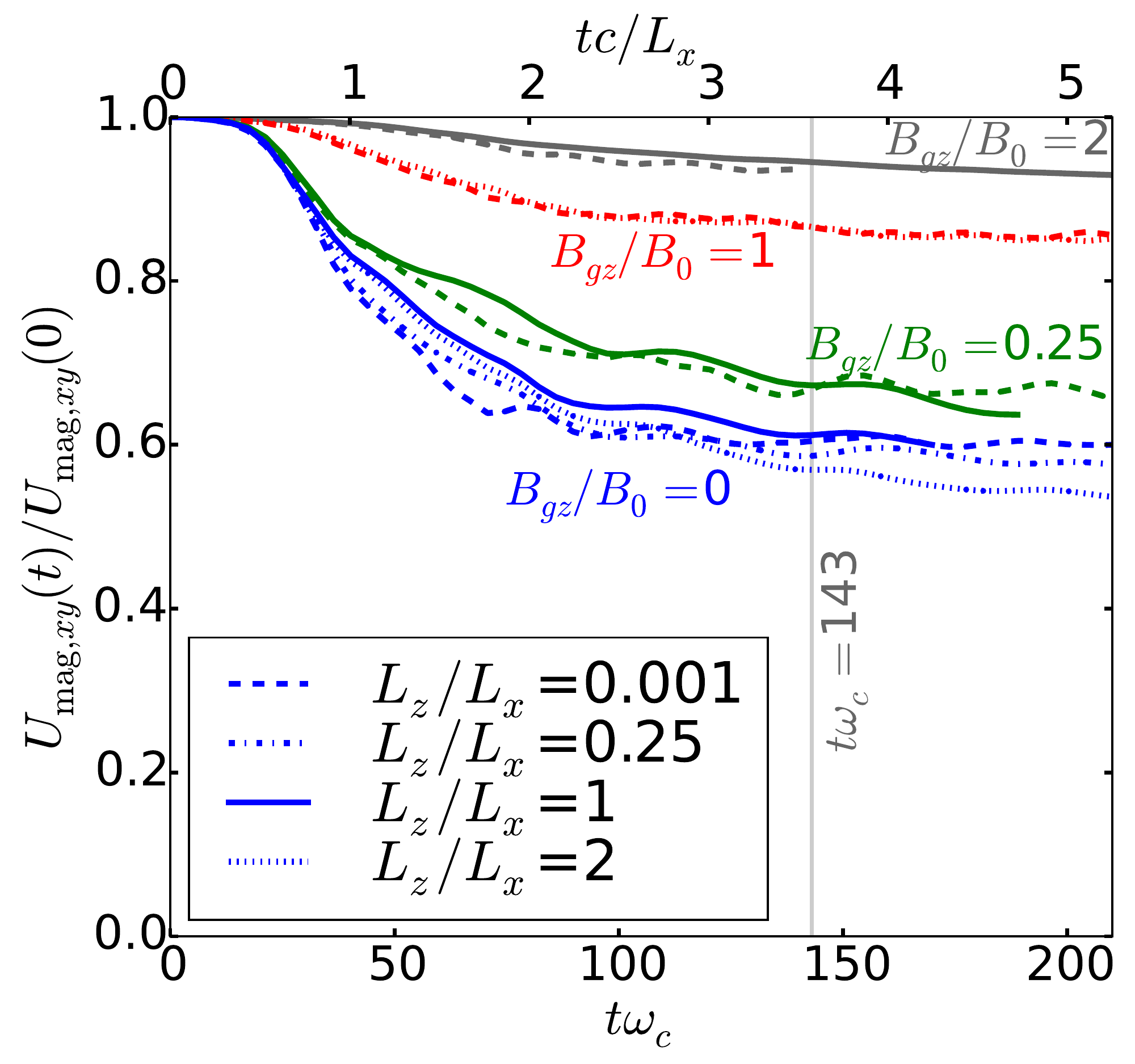}%
\caption{\small
(Left) The electron current density $\|\mathbf{J}\|$ at an early time $t\omega_c=23$ shows both tearing and kinking of the current layer in a simulation with $L_x=L_z=64 \rho_c$ and $B_{gz}=0$ (the image shows a narrow range in $y$ around the layer). 
(Right) 
Despite the layer kinking, the dissipation of transverse magnetic energy versus time is similar across a range of $L_z/L_x$, but greatly reduced by increasing $B_{gz}$.
\label{fig:kinking}
\label{fig:BtVsTime}
}
\end{center}
\end{figure}

In agreement with previous 3D studies \citep{Zenitani_Hoshino-2008,Kagan_etal-2013,Cerutti_etal-2014a}, we find that reconnection proceeds through rapid growth of tearing (plasmoid) and RDKI instabilities, quickly disrupting the initial current sheet and generating a complex, dynamic hierarchy of interacting flux ropes.
Although RDKI has clearly-visible effects 
(e.g., current-layer kinking in the $yz$ plane, Fig.~\ref{fig:kinking}, left),
magnetic dissipation is similar in 2D and~3D.
Figure~\ref{fig:BtVsTime} (right) shows the evolution of transverse magnetic energy, 
$U_{{\rm mag},xy}\equiv \int (B_x^2+B_y^2)/8\pi\, dV$, for different values of $B_{gz}$ and~$L_z/L_x$.
This evolution is typical for a closed system: reconnection ramps up gradually and continues until the available free magnetic energy is exhausted.
The released $U_{{\rm mag},xy}$ is converted mostly into particle kinetic energy, with some into 
$U_{{\rm mag},z}\equiv \int B_z^2/8\pi \, dV$ and a small amount into the electric field (Table~\ref{tab:1}).

Overall, energy dissipation differs little between 2D and $L_z/L_x\gtrsim 1$;
however, 3D simulations with weak guide field ($B_{gz}\lesssim B_0/4$) dissipate slightly more energy than their 2D counterparts by accessing final relaxed states that are nonuniform in~$z$,  e.g., slightly kinked flux ropes with lower magnetic energy than the straight (translationally-symmetric in $z$) flux ropes that are the only possibility in~2D.
In 2D or nearly~2D runs with $B_{gz}=0$, $U_{{\rm mag},xy}$ fell by $\sim$40\% from initial to final state, while decreasing by 46\% for $L_z/L_x=2$ (Fig.~\ref{fig:BtVsTime}, right).
A modest $B_{gz}\gtrsim 0.5$, however, stabilizes the kink (for the explored range of~$L_z/L_x \leq 2$), resulting in nearly identical 2D and 3D energy budgets.

While $L_z/L_x$ has little effect, we find that $B_{gz}$ significantly influences all aspects of reconnection. 
As expected, a strong guide field ($B_{gz} \gtrsim B_0$) suppresses RDKI \citep{Zenitani_Hoshino-2007, Zenitani_Hoshino-2008}. 
In addition, it slows down reconnection and reduces the total dissipated magnetic energy, equally in 2D and~3D. 
We attribute these effects to the guide field's effective inertia and pressure, respectively \citep[see also][]{Dahlin_etal-2016}; 
since the guide field is approximately frozen into the plasma,
its magnetic enthalpy $h_{{\rm mag},z} \equiv B_{gz}^2/4\pi$ lowers the effective $\sigma_h$-parameter,
\begin{eqnarray} \label{eq:sigmaEff}
 \sigma_{h,\rm eff} &\equiv & B_0^2/4\pi(h + h_{{\rm mag},z}) 
    \;=\; B_0^2/(4\pi h + B_{gz}^2)
,\end{eqnarray} 
which governs the relevant transverse Alfv\'en speed, 
$V_A = c\, \sigma_{h,\rm eff}^{1/2}/(1+\sigma_{h,\rm eff})^{1/2} = c\, B_0/(4\pi h + B_{0}^2 + B_{gz}^2)^{1/2}$.
In the high-$\sigma_h$ regime, $B_0^2 \gg 4\pi h$, $V_A$ falls to $V_A/c \approx B_0/(B_0^2 + B_{gz}^2)^{1/2}$, 
becoming subrelativistic for strong guide field ($B_{gz} \gg B_0$),
thereby reducing the reconnection rate, $E_{\rm rec} \sim 0.1 B_0 V_A /c$, to $\sim 0.1 B_0^2/B_{gz}$. 

In addition, since a strong $B_{gz}$ makes the plasma less compressible, it also affects the final relaxed state \citep{Uzdensky_etal-1996}.  
In particular, the tension force of reconnected field lines that contracts plasmoids must perform work against the combined plasma and guide-field pressure;  
a strong $B_{gz}$ resists compression and makes the plasmoids stiffer, reducing the work that can be done and hence the overall amount of $U_{{\rm mag},xy}$ released by reconnection.
Moreover, since some of the released $U_{{\rm mag},xy}$ is spent compressing~$B_z$, the fraction of the energy that goes to particles is also reduced (Table~\ref{tab:1}).

\begin{table}
\caption{
\label{tab:1}
Guide field inhibits magnetic energy dissipation and particle acceleration: for each $B_{gz}$ (for both 2D and $L_z=L_x$), we show $\sigma_{h,\rm eff}$ and
the measured change in $U_{{\rm mag},xy}$, $U_{{\rm mag},z}$,
and particle kinetic energy (normalized to initial $U_{{\rm mag},xy}$), as well as the particle energy spectral index $p$ (the ``error'' 
expresses the fitting uncertainty within a single simulation).
}
\begin{tabular}{llcrrrc}
  $B_{gz}/B_0$ & $\sigma_{h,\rm eff}$ & $L_z/L_x$ &
     $\frac{\displaystyle \Delta U_{{\rm mag},xy}}{\displaystyle U_{{\rm mag},xy}(0)}$ & 
     $\frac{\displaystyle \Delta U_{{\rm mag},z}}{\displaystyle U_{{\rm mag},xy}(0)}$ &
     $\frac{\displaystyle \Delta {\rm KE}}{\displaystyle U_{{\rm mag},xy}(0)}$ 
     & $p$ \\
  \hline
  0 & 25 & 2D & $-$39\% & $\lesssim$0.02\% & 39\% & 1.9$\pm$0.1 \\
  0 & 25 & 1 & $-$46\% & $\lesssim$0.3\% & 46\% & 2.0$\pm$0.1 \\
  0.25 & 9.8 & 2D & $-$34\% & 4\% & 30\% & 2.1$\pm$0.1 \\
  0.25 & 9.8 & 1 & $-$36\% & 5\% & 31\% & 2.1$\pm$0.1 \\
  0.5 & 3.4 & 2D & $-$26\% & 6\% & 19\% & 2.1$\pm$0.1\\
  0.5 & 3.4 & 1 & $-$26\% & 5\% & 20\% & 2.2$\pm$0.1\\
  1  & 0.96 & 2D & $-$15\% & 4\% & 11\% & 3.0$\pm$0.5 \\
  1 & 0.96 & 1 & $-$15\% & 4\% & 11\% & 2.7$\pm$0.2 \\
  2 & 0.25 & 2D & $-$8\% & 1\% & 8\% & 3.3$\pm$0.5 \\
  2 & 0.25 & 1 & $-$8\% & 1\% & 8\% & 3.3$\pm$0.3 \\
\end{tabular}
\end{table}

\begin{figure}[h]
\begin{center}
\includegraphics[width=8.cm,trim=0 0 0 0]{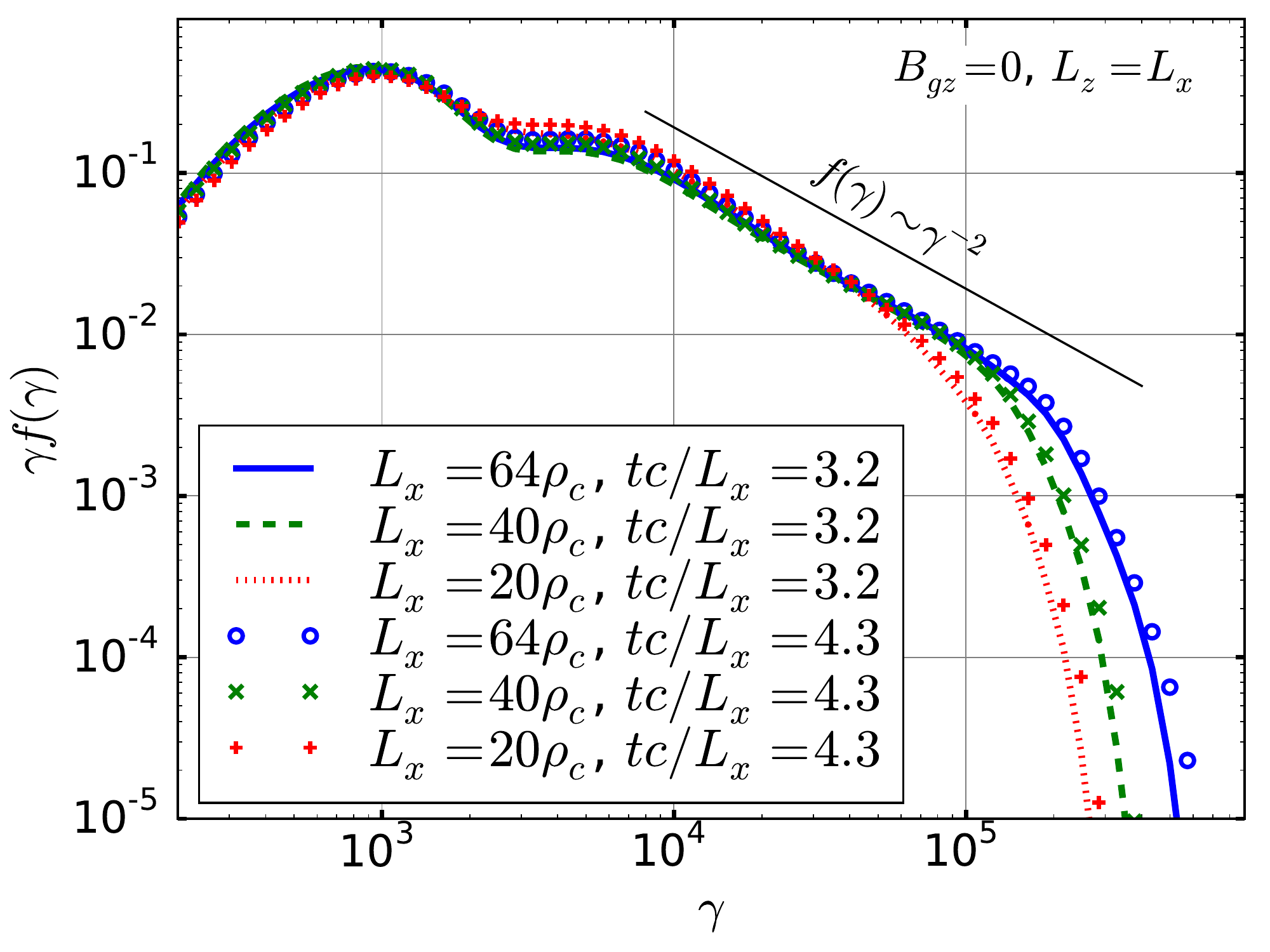}%
\caption{\small
Compensated energy spectra $\gamma f(\gamma)$,  for $B_{gz}=0$ and $L_z=L_x$, saturate at late times; 
the system size $L_x$ has little effect on the power-law slope, but affects the high-energy cutoff.
\label{fig:spectraBz0vsLx}
}
\end{center}
\end{figure}

Most of the released $U_{{\rm mag},xy}$ goes to particles, producing a nonthermal power-law spectrum in both 2D and~3D.
We will show particle energy spectra $f(\gamma)=dN/d\gamma$ at late times, when reconnection has ceased and the spectra stabilize, as illustrated in
Fig.~\ref{fig:spectraBz0vsLx}, which shows nearly identical spectra from times $t=3.2$ and~$4.3\: L_x/c$ for simulations with $B_{gz}=0$ and $L_z=L_x$ (for three different~$L_x$).
For most of our large-system ($L_x=40\rho_c$) runs, the displayed time of $t=143/\omega_c=3.6 L_x/c$ is thus sufficiently late to capture final particle distributions.%
\footnote{
Magnetic dissipation is essentially finished by $t = 143/\omega_c$ in most cases (Fig.~\ref{fig:BtVsTime}, right), except for $B_{gz}=2$, when the spectra evolved until $t \approx 186/\omega_c=4.7L_x/c$.
}

Our simulations reach sufficiently large $L_x$ so that the spectral slopes become nearly independent of system size.  
This is illustrated in Fig.~\ref{fig:spectraBz0vsLx}, which shows $f(\gamma)$ for runs with $L_x/\rho_c\in $\{20, 40, 64\}, $L_z/L_x=1$, and $B_{gz}=0$.
While the power law for $L_x/\rho_c=20$ ($p\approx 2.3$) is slightly  steeper than for $L_x/\rho_c=40$ and~64, the latter two runs have essentially the same $p\approx 2.0$.
Similarly, we find for other $B_{gz}$ and $L_z$ that $p$ converges with $L_x$, within measurement error.

However, the high-energy cutoff $\gamma_c$ of the nonthermal power law still increases with system size (Fig.~\ref{fig:spectraBz0vsLx}). 
This is expected from our previous cutoff investigation in large 2D systems \citep{Werner_etal-2016} employing a similar, but not identical setup, 
which showed that, while $p$ converges at modest~$L_x$, similar to the present study, the rise of $\gamma_c$ with $L_x$ finally saturates only at $L_x/\sigma \rho_0 \sim 200$, well above the sizes ($L_x/\sigma \rho_0=2L_x/\rho_c = $80--128) accessible to our present 3D simulations.
We thus focus here on $p$ and leave determination of $\gamma_c$ to future studies.
Nevertheless, we note that our simulations achieve cutoff energies (in both 2D and 3D) around $\gamma_c \sim 4$--$8 \sigma$, comparable (despite the different setup) to the asymptotic, large-$L_x$ cutoff found in \citep{Werner_etal-2016}, and to those in our 2D study of relativistic electron-ion reconnection \citep{Werner_etal-2017submitted}.
This indicates that 3D reconnection can indeed accelerate particles to about the same energies as~2D.

\begin{figure}
\begin{center}
\includegraphics[height=4.4cm,trim=0 0 0 0]{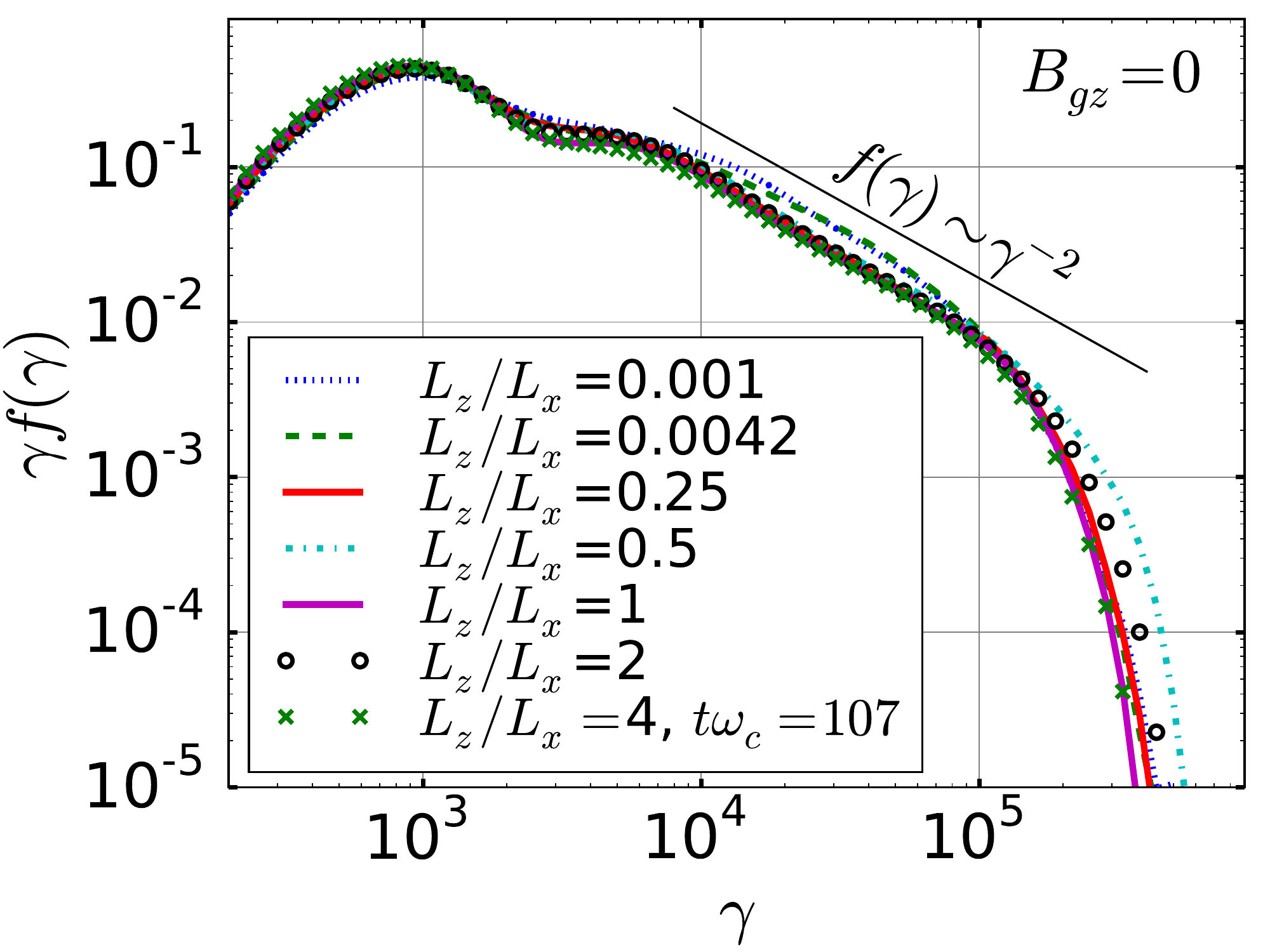}%
\includegraphics[height=4.4cm,trim=36mm 0 0 0,clip]{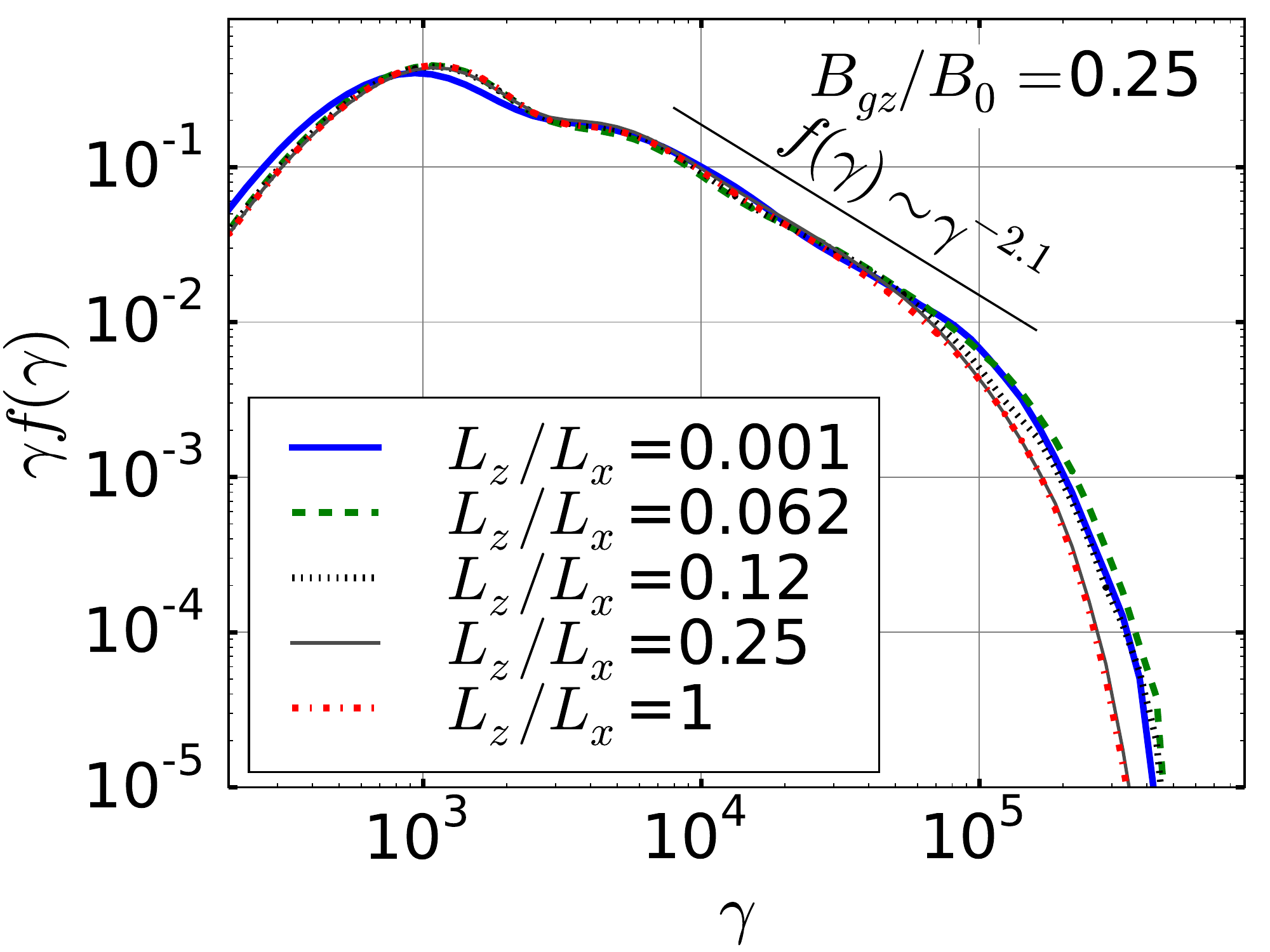}%
\includegraphics[height=4.4cm,trim=36mm 0 0 0,clip]{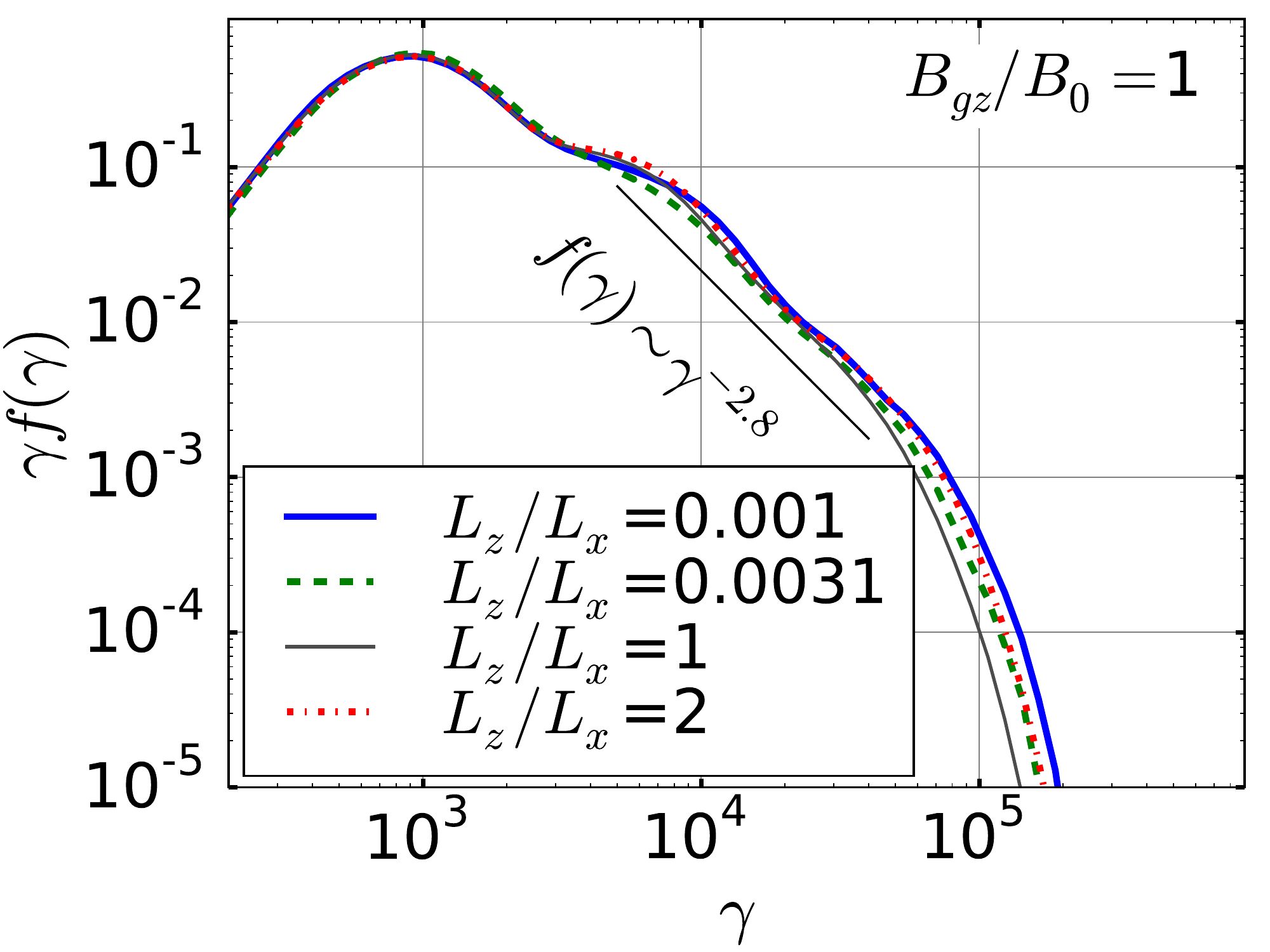}%
\caption{
Particle energy spectra for different aspect ratios $L_z/L_x$, reflecting the importance of 3D effects, for $B_{gz}/B_0\in \{0, 0.25, 1\}$.
The spectra are essentially independent of the aspect ratio.
The spectra are shown at 
$t = 143 \omega_c^{-1} = 3.6 L_x/c$, except for $L_z/L_x=4$, $B_{gz}=0$, shown at
$t = 107 \omega_c^{-1} = 2.7 L_x/c$ when that simulation ended prematurely.
\label{fig:spectraVsLz}
}
\end{center}
\end{figure}

We now characterize the effects of three-dimensionality and guide field on  NTPA, comparing simulations with the same size $L_x=40\rho_c=80\sigma \rho_0$ but different $L_z/L_x$ and~$B_{gz}$.
Our main finding is that, despite fundamental differences in~2D and 3D dynamics (Fig.~\ref{fig:kinking}, left), particle acceleration is nearly unchanged by 3D effects, regardless of guide field (Fig.~\ref{fig:spectraVsLz}). Importantly, 3D relativistic reconnection with weak or no guide field is an efficient particle accelerator, 
consistent with single simulations run by 
\cite{Sironi_Spitkovsky-2014} and \cite{Guo_etal-2014,Guo_etal-2015},
but contradicting \cite{Zenitani_Hoshino-2007, Zenitani_Hoshino-2008}.%
\footnote{
{\citet{Dahlin_etal-2015} nonrelativistic electron-ion reconnection simulations without initial perturbation found more efficient NTPA in 3D due to weaker particle trapping allowing multiple re-acceleration of particles.}
}
We see some small differences in $p$ between 2D and 3D in the $B_{gz}=0$ case (Fig.~\ref{fig:spectraVsLz}, left); however, for finite guide field, $B_{gz}/B_0=0.25$ and~1 (Fig.~\ref{fig:spectraVsLz} middle and right), the 2D and 3D spectra are almost identical, probably because strong~$B_{gz}$ suppresses dynamical variation in~$z$.

\begin{figure}[t]
\begin{center}
\includegraphics[width=7.4cm,trim=0 0 0 0]{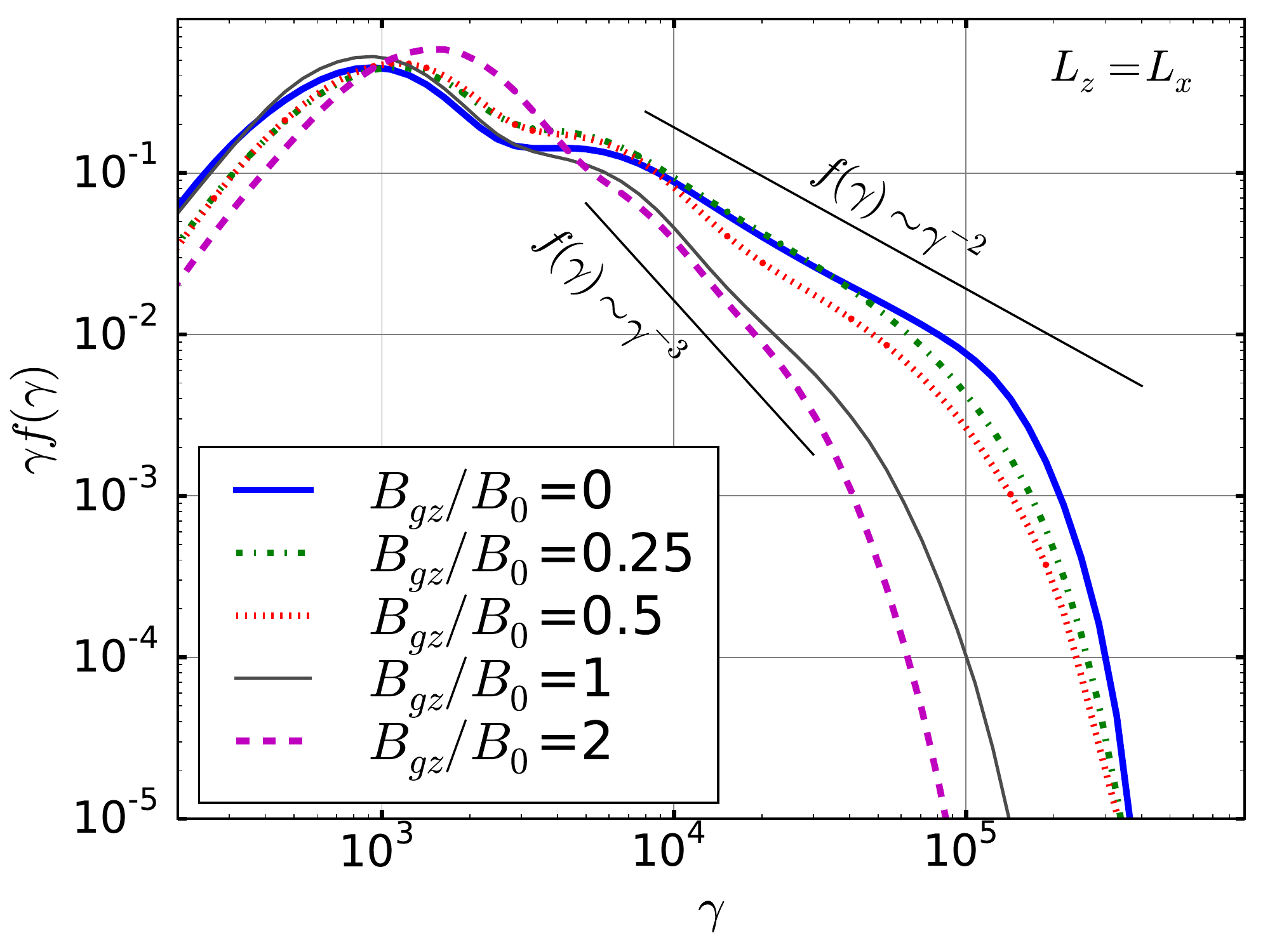}%
\includegraphics[width=7.6cm,trim=0 0 0 0]{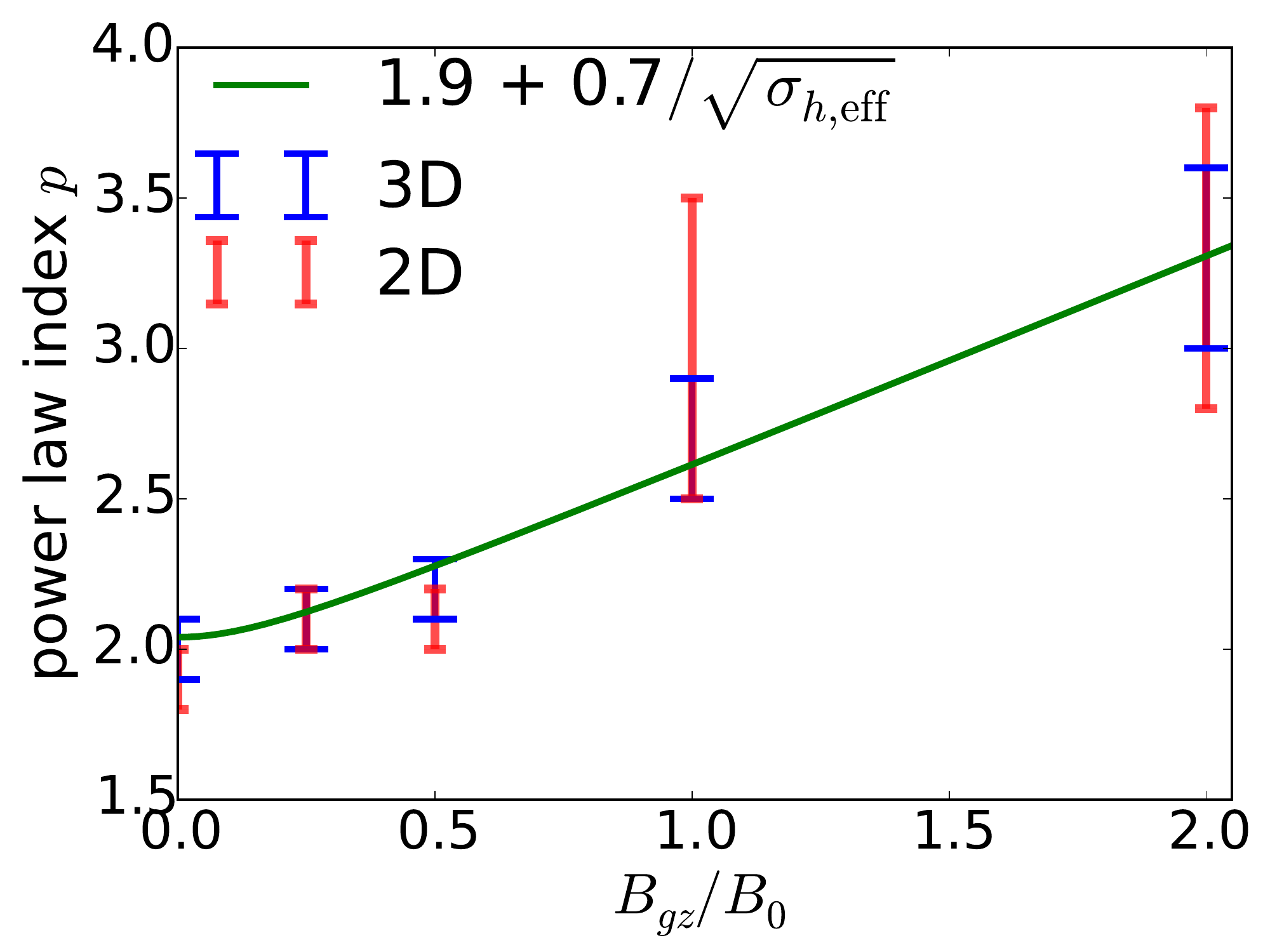}%
\caption{
(Left) Strong $B_{gz}$ hinders particle acceleration,
as shown by the particle spectra from simulations with $L_z=L_x$ and varying~$B_{gz}$.
\label{fig:spectraVsBz}
(Right) The spectral slopes are similar in 2D and~3D, but steepen significantly with strong guide field [Eq.~(\ref{eq:sigmaEff})].
The range of $p$ indicates variation within a single simulation.
\label{fig:pFits}
}
\end{center}
\end{figure}

Finally, we find that a sufficiently strong $B_{gz}$ hinders NTPA in 2D and~3D, reducing the number and total energy of accelerated particles, and steepening the power law (see Fig.~\ref{fig:spectraVsBz}, left).
The effects of three-dimensionality and guide field are summarized in Fig.~\ref{fig:pFits} (right), comparing 2D and 3D ($L_z=L_x$) simulations.
We conjecture that the lower reconnection rate (hence weaker accelerating electric field) and the smaller available energy budget (Table~\ref{tab:1}) both contribute to the relative inefficiency and limited range of NTPA  for strong guide fields, $B_{gz} \gtrsim B_0$. 
We further propose that~$p$ steepens with $B_{gz}$ because $\sigma_{h,\rm eff}\sim B_0^2/B_{gz}^2$ decreases [Eq.~(\ref{eq:sigmaEff})], and lowering $\sigma_h$ has been seen to increase~$p$ \citep[in $B_{gz}=0$ 2D studies, ][]{Sironi_Spitkovsky-2014, Guo_etal-2014, Werner_etal-2016}.
Intriguingly, while it does not yield $p\sim 1$ for $\sigma_{h,\rm eff}\rightarrow \infty$ as those 2D studies suggest, the empirical fit $p\approx 1.9+0.7\sigma_{h,\rm eff}^{-1/2}$, found for $B_{gz}=0$ semirelativistic electron-ion reconnection \citep{Werner_etal-2017submitted}, captures the $B_{gz}$-dependence found here.


\section{Conclusions} 
\label{sec-conclusions}

This Letter presents a systematic first-principles confirmation via PIC simulation that 3D relativistic reconnection in pair plasmas can robustly and efficiently drive nonthermal particle acceleration (NTPA), yielding unambiguous power-law particle distributions despite the presence of RDKI.
In addition, the strong influence of guide field on the NTPA power-law index $p$ is described through the dependence of $p$ \citep[see ][]{Werner_etal-2017submitted} on the effective hot magnetization $\sigma_{h,\rm eff}$ including the enthalpy of the guide field.
This study thus resolves the long-standing controversy regarding the effects of 3D RDKI structures and guide field on NTPA \citep{Jaroschek_etal-2004a,Zenitani_Hoshino-2007,Zenitani_Hoshino-2008,Sironi_Spitkovsky-2014,Guo_etal-2014}. 
Importantly, our results show that 2D studies are in fact pertinent to 3D reconnection; intriguingly, however, the poorer particle trapping within 3D flux ropes \citep{Dahlin_etal-2015} may allow 3D NTPA to extend beyond the energy cutoff in 2D reconnection \citep{Werner_etal-2016}.
Our findings lend strong support to reconnection-based models of, e.g., rapid $\gamma$-ray flares in the Crab PWN and AGN/blazar jets,  
prompt GRB emission, and hard-X-ray emission in accreting BH binaries (cf.~\S\ref{sec-intro}). 
Furthermore, these results lay the foundation for characterizing NTPA in realistic 3D reconnection as a function of ambient plasma, helping to diagnose plasma conditions in remote astrophysical systems using observed radiation spectra.


\acknowledgments
We thank Vladimir Zhdankin and Mitch Begelman for helpful discussions.
This work was supported by grants DE-SC0008409, DE-SC0008655 (DOE); 
NNX12AP17G, NNX16AB28G (NASA); and AST-1411879 (NSF).
D.A.U. gratefully acknowledges the hospitality of the Institute for Advanced Study and support from the Ambrose Monell Foundation.
Simulations were possible thanks to an INCITE award at the Argonne Leadership Computing Facility, a DOE User Facility supported under Contract DE-AC02-06CH11357.  Some 2D simulations were run using XSEDE resources \citep{XSEDE2014} supported by NSF grant ACI-1548562, as well as resources provided by the NASA High-End Computing Program through the NASA Advanced Supercomputing Division at Ames Research Center.

\bibliographystyle{aasjournal}

\begin{thebibliography}{}
\expandafter\ifx\csname natexlab\endcsname\relax\def\natexlab#1{#1}\fi
\providecommand{\url}[1]{\href{#1}{#1}}

\bibitem[{Arka \& Dubus(2013)}]{Arka_Dubus-2013}
Arka, I., \& Dubus, G. 2013, A\& A, 550, A101

\bibitem[{Beloborodov(2017)}]{Beloborodov-2017}
Beloborodov, A.~M. 2017, arXiv preprint arXiv:1701.02847

\bibitem[{Bessho \& Bhattacharjee(2007)}]{Bessho_Bhattacharjee-2007}
Bessho, N., \& Bhattacharjee, A. 2007, Physics of Plasmas, 14, 056503

\bibitem[{Bhattacharjee {et~al.}(2009)Bhattacharjee, Huang, Yang, \&
  Rogers}]{Bhattacharjee_etal-2009}
Bhattacharjee, A., Huang, Y.-M., Yang, H., \& Rogers, B. 2009, Physics of
  Plasmas, 16, 112102

\bibitem[{Cerutti {et~al.}(2015)Cerutti, Philippov, Parfrey, \&
  Spitkovsky}]{Cerutti_etal-2015}
Cerutti, B., Philippov, A., Parfrey, K., \& Spitkovsky, A. 2015, MNRAS, 448,
  606

\bibitem[{{Cerutti} {et~al.}(2012{\natexlab{a}}){Cerutti}, {Uzdensky}, \&
  {Begelman}}]{Cerutti_etal-2012a}
{Cerutti}, B., {Uzdensky}, D.~A., \& {Begelman}, M.~C. 2012{\natexlab{a}}, ApJ,
  746, 148

\bibitem[{{Cerutti} {et~al.}(2012{\natexlab{b}}){Cerutti}, {Werner},
  {Uzdensky}, \& {Begelman}}]{Cerutti_etal-2012b}
{Cerutti}, B., {Werner}, G.~R., {Uzdensky}, D.~A., \& {Begelman}, M.~C.
  2012{\natexlab{b}}, ApJ Lett., 754, L33

\bibitem[{{Cerutti} {et~al.}(2013){Cerutti}, {Werner}, {Uzdensky}, \&
  {Begelman}}]{Cerutti_etal-2013}
---. 2013, ApJ, 770, 147

\bibitem[{{Cerutti} {et~al.}(2014{\natexlab{a}}){Cerutti}, {Werner},
  {Uzdensky}, \& {Begelman}}]{Cerutti_etal-2014b}
---. 2014{\natexlab{a}}, ApJ, 782, 104

\bibitem[{{Cerutti} {et~al.}(2014{\natexlab{b}}){Cerutti}, {Werner},
  {Uzdensky}, \& {Begelman}}]{Cerutti_etal-2014a}
---. 2014{\natexlab{b}}, Phys. Plasmas, 21, 056501

\bibitem[{{Coroniti}(1990)}]{Coroniti-1990}
{Coroniti}, F.~V. 1990, ApJ, 349, 538

\bibitem[{Dahlin {et~al.}(2015)Dahlin, Drake, \& Swisdak}]{Dahlin_etal-2015}
Dahlin, J., Drake, J., \& Swisdak, M. 2015, Phys. Plasmas, 22, 100704

\bibitem[{Dahlin {et~al.}(2016)Dahlin, Drake, \& Swisdak}]{Dahlin_etal-2016}
---. 2016, Phys. Plasmas, 23, 120704

\bibitem[{{Drenkhahn} \& {Spruit}(2002)}]{Drenkhahn_Spruit-2002}
{Drenkhahn}, G., \& {Spruit}, H.~C. 2002, A\& A, 391, 1141

\bibitem[{Giannios(2008)}]{Giannios-2008}
Giannios, D. 2008, A\& A, 480, 305

\bibitem[{{Giannios} {et~al.}(2009){Giannios}, {Uzdensky}, \&
  {Begelman}}]{Giannios_etal-2009}
{Giannios}, D., {Uzdensky}, D.~A., \& {Begelman}, M.~C. 2009, MNRAS, 395, L29

\bibitem[{Guo {et~al.}(2014)Guo, Li, Daughton, \& Liu}]{Guo_etal-2014}
Guo, F., Li, H., Daughton, W., \& Liu, Y.-H. 2014, Phys. Rev. Lett., 113,
  155005

\bibitem[{Guo {et~al.}(2015)Guo, Liu, Daughton, \& Li}]{Guo_etal-2015}
Guo, F., Liu, Y.-H., Daughton, W., \& Li, H. 2015, ApJ, 806, 167

\bibitem[{{Jaroschek} {et~al.}(2004){Jaroschek}, {Treumann}, {Lesch}, \&
  {Scholer}}]{Jaroschek_etal-2004a}
{Jaroschek}, C.~H., {Treumann}, R.~A., {Lesch}, H., \& {Scholer}, M. 2004,
  Physics of Plasmas, 11, 1151

\bibitem[{{Kagan} {et~al.}(2013){Kagan}, {Milosavljevi{\'c}}, \&
  {Spitkovsky}}]{Kagan_etal-2013}
{Kagan}, D., {Milosavljevi{\'c}}, M., \& {Spitkovsky}, A. 2013, ApJ, 774, 41

\bibitem[{{Kirk} \& {Skj{\ae}raasen}(2003)}]{Kirk_Skjaeraasen-2003}
{Kirk}, J.~G., \& {Skj{\ae}raasen}, O. 2003, ApJ, 591, 366

\bibitem[{Langdon(1970)}]{Langdon-1970}
Langdon, A.~B. 1970, J. Comput. Phys., 6, 247

\bibitem[{{Liu} {et~al.}(2011){Liu}, {Li}, {Yin}, {Albright}, {Bowers}, \&
  {Liang}}]{Liu_etal-2011}
{Liu}, W., {Li}, H., {Yin}, L., {et~al.} 2011, Physics of Plasmas, 18, 052105

\bibitem[{Liu {et~al.}(2015)Liu, Guo, Daughton, Li, \& Hesse}]{Liu_etal-2015}
Liu, Y.-H., Guo, F., Daughton, W., Li, H., \& Hesse, M. 2015, Phys. Rev. Lett.,
  114, 095002

\bibitem[{Loureiro {et~al.}(2012)Loureiro, Samtaney, Schekochihin, \&
  Uzdensky}]{Loureiro_etal-2012}
Loureiro, N., Samtaney, R., Schekochihin, A., \& Uzdensky, D. 2012, Physics of
  Plasmas, 19, 042303

\bibitem[{{Lyubarsky} \& {Liverts}(2008)}]{Lyubarsky_Liverts-2008}
{Lyubarsky}, Y., \& {Liverts}, M. 2008, ApJ, 682, 1436

\bibitem[{{Lyubarsky}(1996)}]{Lyubarsky-1996}
{Lyubarsky}, Y.~E. 1996, A\& A, 311, 172

\bibitem[{Lyutikov(2003)}]{Lyutikov-2003}
Lyutikov, M. 2003, MNRAS, 346, 540

\bibitem[{Lyutikov(2006)}]{Lyutikov-2006}
---. 2006, MNRAS, 367, 1594

\bibitem[{{McKinney} \& {Uzdensky}(2012)}]{McKinney_Uzdensky-2012}
{McKinney}, J.~C., \& {Uzdensky}, D.~A. 2012, MNRAS, 419, 573

\bibitem[{Melzani {et~al.}(2014)Melzani, Walder, Folini, Winisdoerffer, \&
  Favre}]{Melzani_etal-2014b}
Melzani, M., Walder, R., Folini, D., Winisdoerffer, C., \& Favre, J.~M. 2014,
  A\& A, 570, A112

\bibitem[{{Nalewajko} {et~al.}(2011){Nalewajko}, {Giannios}, {Begelman},
  {Uzdensky}, \& {Sikora}}]{Nalewajko_etal-2011}
{Nalewajko}, K., {Giannios}, D., {Begelman}, M.~C., {Uzdensky}, D.~A., \&
  {Sikora}, M. 2011, MNRAS, 413, 333

\bibitem[{{Sironi} {et~al.}(2016){Sironi}, {Giannios}, \&
  {Petropoulou}}]{Sironi_etal-2016}
{Sironi}, L., {Giannios}, D., \& {Petropoulou}, M. 2016, \mnras, 462, 48

\bibitem[{{Sironi} {et~al.}(2015){Sironi}, {Petropoulou}, \&
  {Giannios}}]{Sironi_etal-2015}
{Sironi}, L., {Petropoulou}, M., \& {Giannios}, D. 2015, \mnras, 450, 183

\bibitem[{{Sironi} \& {Spitkovsky}(2014)}]{Sironi_Spitkovsky-2014}
{Sironi}, L., \& {Spitkovsky}, A. 2014, ApJ Lett., 783, L21

\bibitem[{Towns {et~al.}(2014)Towns, Cockerill, Dahan, Foster, Gaither,
  Grimshaw, Hazlewood, Lathrop, Lifka, Peterson, {et~al.}}]{XSEDE2014}
Towns, J., Cockerill, T., Dahan, M., {et~al.} 2014, Computing in Science \&
  Engineering, 16, 62

\bibitem[{Uzdensky(2011)}]{Uzdensky-2011}
Uzdensky, D.~A. 2011, Space science reviews, 160, 45

\bibitem[{{Uzdensky} {et~al.}(2011){Uzdensky}, {Cerutti}, \&
  {Begelman}}]{Uzdensky_etal-2011}
{Uzdensky}, D.~A., {Cerutti}, B., \& {Begelman}, M.~C. 2011, ApJ Lett., 737,
  L40

\bibitem[{Uzdensky {et~al.}(1996)Uzdensky, Kulsrud, \&
  Yamada}]{Uzdensky_etal-1996}
Uzdensky, D.~A., Kulsrud, R.~M., \& Yamada, M. 1996, Phys. Plasmas, 3, 1220

\bibitem[{Uzdensky {et~al.}(2010)Uzdensky, Loureiro, \&
  Schekochihin}]{Uzdensky_etal-2010}
Uzdensky, D.~A., Loureiro, N.~F., \& Schekochihin, A.~A. 2010, Phys. Rev.
  Lett., 105, 235002

\bibitem[{{Uzdensky} \& {Spitkovsky}(2014)}]{Uzdensky_Spitkovsky-2014}
{Uzdensky}, D.~A., \& {Spitkovsky}, A. 2014, ApJ, 780, 3

\bibitem[{{Werner} {et~al.}(2017){Werner}, Uzdensky, {Begelman}, {Cerutti}, \&
  {Nalewajko}}]{Werner_etal-2017submitted}
{Werner}, G.~R., Uzdensky, D.~A., {Begelman}, M.~C., {Cerutti}, B., \&
  {Nalewajko}, K. 2017, submitted, arXiv:1612.04493

\bibitem[{{Werner} {et~al.}(2016){Werner}, Uzdensky, {Cerutti}, {Nalewajko}, \&
  {Begelman}}]{Werner_etal-2016}
{Werner}, G.~R., Uzdensky, D.~A., {Cerutti}, B., {Nalewajko}, K., \&
  {Begelman}, M.~C. 2016, ApJ Lett., 816, L8

\bibitem[{{Zenitani} \& {Hoshino}(2001)}]{Zenitani_Hoshino-2001}
{Zenitani}, S., \& {Hoshino}, M. 2001, ApJ Lett., 562, L63

\bibitem[{{Zenitani} \& {Hoshino}(2005)}]{Zenitani_Hoshino-2005}
---. 2005, ApJ Lett., 618, L111

\bibitem[{{Zenitani} \& {Hoshino}(2007)}]{Zenitani_Hoshino-2007}
---. 2007, ApJ, 670, 702

\bibitem[{{Zenitani} \& {Hoshino}(2008)}]{Zenitani_Hoshino-2008}
---. 2008, ApJ, 677, 530

\end{thebibliography}

\end{document}